
\def\PRD{Phys. Rev. }\def\MPLA{Mod. Phys. Lett. }\def\NPB{Nucl. Phys. }
\def\PRL{Phys. Rev. Lett. }
\def\sect{\bigbreak\noindent\bf}\def\ssect{\noindent\sl}
\def\af{asymptotically flat }\def\ads{anti-de Sitter }
\def\bh{black hole }\def\st{spacetime }\def\el{extremal limit }
\def\ld{linear dilaton }\def\bg{background }\def\pd{direct product }
\def\Sch{Schwarzschild }\def\coo{coordinates }\def\gs{ground state }
\def\Mi{Minkowski }
\def\ph{parabolic hyperboloids }
\def\dr{dimensional reduction }\def\bc{boundary conditions }
\def\ds{ds^2=}\def\ef{e^{2(\phi-\phi_0)}}\def\F{F_\mn}\def\gmn{g_{\mu\nu}}
\def\kp{{1+k\over 2}}\def\km{{1-k\over 2}}\def\ct{\tilde\chi}
\def\kf{{1-k\over 1+k}}\def\kk{{2k\over 1+k}}\def\kd{{-{2\over 1+k}}}
\def\rp{r_+}\def\rq{r_-}\def\qy{\left(1+{Q\over y}\right)}
\def\qf{{3\over q}\phi}\def\kg{{1+k\over 1-k}}
\def\lo{\lambda_1^2}\def\lt{\lambda_2^2}\def\efp{e^{2(\phi -{q\over 3}\psi)}}
\def\kq{{1-k\over 1+k}}\def\kr{{1+k\over 1-k}}\def\mn{{\mu\nu}}
\def\dd{\partial_+}\def\dm{\partial_-}
\def\xm{x^-}\def\xn{x^+}\def\ma{\left(\xm - \xn\right)}
\def\la{\lambda_2}
\def\ks{{k\over 1-k}}
\def\kl{{k-1\over 2}}
\def\ka{{k-1\over 1+k}}
\magnification=1200
\line {December 1993 \hfil  INFN-CA-20-93}
 \vfill
\centerline {\bf CLASSICAL AND SEMICLASSICAL PROPERTIES OF EXTREMAL  }
\centerline {\bf BLACK HOLES WITH DILATON AND MODULUS FIELDS}
\vskip 1.2in
\centerline {\bf M. Cadoni}
\centerline {\bf S. Mignemi}
\vskip .2in

\centerline {Dipartimento di Scienze Fisiche, Universit\`a di Cagliari}
\centerline {and Istituto Nazionale di Fisica Nucleare, Sezione di Cagliari}
\centerline {Via Ada Negri 18, I-09127 Cagliari, Italy }\vfill
\centerline {\bf ABSTRACT}

We discuss both classical and semiclassical properties of extremal black holes
in  theories where the dilaton and a modulus field are present.
We find that the corresponding 2-dim geometry is asymptotically anti-de
Sitter rather then asymptotically flat as in the purely dilatonic case.
This fact has many  important consequences, which we analyze at length,
both for the classical behaviour and for the thermodynamical properties
of the black hole. We also study the Hawking evaporation process in the
semiclassical
approximation. The calculations strongly indicates the emergence of a stable
ground state  as the end point of the process. Some comments are made about the
relevance of our results for the problem of information loss in black hole
physics.

\eject

{\sect 1. Introduction}
\smallskip
String inspired \bh models have been extensively studied in recent years.
A first motivation for this interest is the fact that exact charged solutions
of the 4-dim lowest order effective action of string theory have been found
in a variety of cases [1-3].

These solutions exhibit peculiar thermodynamical properties, which are
different from those of the ordinary Einstein theory black holes [4].
In particular, the extremal limit,  where the values of the mass and of the
charge are such that the \bh is on the edge of becoming a naked singularity,
presents a behaviour which resembles that of elementary particles [5].
Even if some of these features are spoiled by going to higher order in
perturbation theory [6], these models still present a relevant theoretical
interest. The extremal limit of a black hole of given charge represents,
in fact, the ground state for its
Hawking evaporation process. The study of the scattering of low-energy
particles
by an extremal \bh should therefore shed some light on the problem of \bh
evaporation and on the fate of the information after the evaporation process
has taken place [7]. In particular, one hopes to clarify the issue whether the
\bh completely disappears or a massive remnant is left.
In this context,  a very useful property of extremal string \bh solutions
is the fact that in proximity of the horizon they split
into a direct product of a 2-dim solution with a 2-sphere of constant
radius [8]. It is therefore possible to
study the properties of the \bh by means of a
reduced 2-dimensional model, which makes the problem much easier to treat [9].
Many papers have in fact been devoted to the investigation of these 2-dim
models [10,11]. If the back-reaction of the gravitational field is neglected,
exact solution of the field equations including one-loop quantum
corrections can be found, which describe the formation of a \bh and the early
stages of its evaporation [9].
Moreover, if the back-reaction is also taken into account, the
behaviour of the black hole during the final stages of the evaporation can be
qualitatively described [10].

The peculiar properties of string black holes are essentially due to
the non-minimal coupling of gravity and gauge fields to a massless scalar
field which is contained in the low-energy spectrum of string, namely the
dilaton. The non-minimality of the coupling permits to circumvent the
no-hair theorems which render essentially unique the \bh solutions of
Einstein theory. It is therefore interesting to consider the other
non-minimally coupled massless scalars which appear in the
low-energy effective action of the string [3,12]. It turns out, in fact, that
modulus fields coming from the compactification of the string to 4-dimensions
can acquire non minimal couplings to the gauge fields of the 4-dim action
owing to string one-loop effects [13].
In a recent paper [3], we have studied a model where these effects are
taken into account and found a class of exact 4-dim solutions,
whose properties are slightly different from those where modulus fields
are decoupled. In this paper we extend the investigation of such solutions to
their extremal limit and study the \bh evaporation by means of a 2-dim
effective model.

The main result of our investigation is that, contrary to the pure dilatonic
case, the solutions of the 2-dim effective theory are not asymptotically flat,
but have as asymptotic a space of constant negative curvature.
This implies  that in this approximation    the 4-dim extremal
\bh is screened by an infinite potential barrier, which prevents it to
irradiate
to the asymptotically flat region. This explains why the temperature
vanishes in the 4-dim theory. We also study the quantum evaporation
of the 2-dim \bh. We find that for a wide range of the parameters
which characterize the model, the Hawking radiation rate is not asymptotically
mass independent as in the case of pure dilatonic 2-dim gravity, but goes
to zero with the mass of the hole. This strongly suggests the emergence of
a stable state at the end point of the evaporation process.

The paper is organized as follows: in section 2 we review the 4-dim solutions
found in [3] and study their extremal limit. In section 3 we examine the
propagation of a perturbation in the throat region of the solution
and show that it is exponentially damped at infinity.
In section 4 we study the 2-dimensional effective theory and discuss its
solutions in different gauges. We also review 2-dim spacetimes of constant
curvature. In section 5 we study  the 2-dimensional theory in the conformal
gauge and in section 6 we analyse the 2-dim  \bh evaporation process using
various methods. We state our conclusions in section 7.

\bigskip
{\sect 2. The solution in 4 dimensions}

{\ssect a) The general solution}
\smallskip
In a recent paper [3], we have studied the magnetic charged black hole
solution of a 4-dim gravitational action obtained by low-energy effective
string theory when moduli of the compactified manifold are taken into
account.

In terms of the metric to which the string couples, which we shall use
in this paper, the action had the form:
$$S=\int\sqrt g\ d^4x\ e^{-2\phi}\left[R+4(\nabla\phi)^2-{2\over
3}(\nabla\psi)
^2-F^2-\efp F^2\right],\eqno(2.1)$$
where $\phi$ is the dilaton field, $\psi$ the compacton and $\F$ the Maxwell
field strength, $q$ being a coupling constant.

Exact charged \bh solutions to this action were found in the case $\psi=\qf +
const$,
which have the form:
$$\ds -\left(1-{\rp\over r}\right)\left(1-{\rq\over r}\right)^k dt^2
+\left(1-{\rp\over r}\right)^{-1}\left(1-{\rq\over r}\right)^{-1} dr^2
+r^2 d\Omega^2,$$
$$\ef=\left(1-{\rq\over r}\right)^{(k-1)/2},\qquad\qquad\F={Q_M\over r^2}
\epsilon_\mn,\eqno(2.2)$$
with
$$k={3-2q^2\over 3+2q^2},\qquad\qquad -1\le k\le 1.\eqno(2.3)$$

The two parameters $r_+$ and $r_-$ are related to the charge $Q$ and
to the mass $M$ of the  \bh by the relations:
$$2M=r_++{k+1\over 2}r_-,\qquad\qquad Q_M^2={1-k\over 4}r_+r_-.\eqno(2.4)$$
In the limit $k=-1$ (i.e. $q\to\infty$), the solution reduces to that found in
in ref. [1] for the case where the compacton field is not taken into account.

The spatial part of the metric (2.2) is actually identical to that of ref.
[1]: it describes an \af region attached to a long tube (the "throat"),
terminated by a horizon. In the extremal limit $\rp=\rq$, the tube becomes
infinitely long.

In our case, however, the full spacetime has a slightly different structure
with respect to ref. [1]: as we shall see, the infinitely long tube in the
extremal limit is replaced by the direct product of a 2-dim negative
curvature \st with a 2-sphere of constant radius. This fact will have
important consequences on the physical properties of the model.

Also of interest are the thermodynamical properties of the black hole
described by (2.2). The temperature is given by
$$T={1\over 4\pi\rp}\left(1-{\rq\over\rp}\right)^\kp.\eqno(2.5)$$
which goes to zero in the extremal limit $\rp=\rq$, while the entropy is
$$S=\pi r_+^2.\eqno(2.6a)$$
One could therefore think of the \el as a non-radiating ground state. This
point of view will be confirmed in the following.

It should be noticed, however, that for the "canonical" metric $\tilde g
=e^{-2\phi}g$, while the expression for the temperature remains the same,
the entropy is given by
$$S=\pi r_+^2\left(1-{\rq\over\rp}\right)^\km,\eqno(2.6b)$$
which also vanishes for $\rp\to\rq$. The discrepance is of course due to the
different definition of the volume element in the two metrics.

\bigskip
{\ssect b) The extremal limit}
\smallskip
In the rest of this paper we shall study the \el $\rp=\rq=Q$ of the
solution (2.2), where $Q={2\over\sqrt{1-k}}Q_M$,
and the ensuing 2-dim effective theory, along the lines of ref. [8].
For this purpose, it is useful to define a new coordinate $\sigma$, such
that $\sigma={\rm arcsh}\sqrt{(r-\rp)\over (\rp-\rq)}$.

In terms of $\sigma$, the metric and the dilaton field (2.2) take the form
$$\ds -4Q^2{\Delta^{k+1}\sinh^2\sigma\cosh^{2k}\sigma\over (\rp+\Delta\sinh^2
\sigma)^{k+1}}dt^2+(\rp+\Delta\sinh^2\sigma)^2(4d\sigma^2+d\Omega^2),$$
$$\ef=\left[{\rp+\Delta\sinh^2\sigma\over\Delta\cosh^2\sigma}\right]^\km,
\eqno(2.7)$$
where $\Delta=\rp-\rq$.

There are several regimes under which the \el can be approached, which
correspond to different solutions of the action (2.1):

1) $\sigma\gg 1$: this limit is reached by taking the \af region and
the throat
fixed and allowing the horizon to move off to infinity as $\Delta\to 0$.
The solution is then:
$$\ds-4Q^2\qy^{-(k+1)}dt^2+\qy^2(dy^2+y^2d\Omega^2),$$
$$\ef=\qy^\km,\qquad\qquad\F={\sqrt{1-k}\over 2} {Q\over (y+Q)^2}
\epsilon_\mn,\eqno(2.8)$$
with $y\ge 0$. The metric is everywhere regular and describes the transition
between an \af \st for $y\to\infty$ and one with topology $H^2\times S^2$ for
$y\to 0$, $H^2$ being 2-dim \ads \st. This solution can therefore be
considered a generalization of the solitonic solutions of string theory
described in [14].

2) $1\gg\sigma\gg\ln (Q/\Delta)$: this limit corresponds to the infinite throat
with \ld and is reached by sending to infinity both the horizon and the \af
region:
$$\ds-4Q^2e^{2(k+1)\sigma}dt^2+Q^2(4d\sigma^2+d\Omega^2),$$
$$\phi=\kl\sigma.\eqno(2.9)$$
At variance with the GHS case, the metric in this limit does not describe an
infinite cylinder, but rather the direct product of 2-dim \ads \st with a
2-sphere of radius $Q$.

3) $\sigma\ll 1$: this limit is obtained by keeping the horizon and $\Delta
^\km e^{-2\phi_0}$ fixed, and letting the \af region go to infinity when
$\Delta
\to 0$ and describes the horizon plus the infinite throat. The solution is
given by:
$$\ds-4Q^2\sinh^2\sigma\cosh^{2k}\sigma\ dt^2+Q^2(4d\sigma^2+d\Omega^2),$$
$$\ef=\left({Q\over\cosh^2\sigma}\right)^\km,\eqno(2.10)$$
and again has the form of a  direct product of a 2-dim solutions with an
horizon at $\sigma =0$ and a 2-sphere. It will therefore be useful in the
discussion of \bh solutions of the 2-dim effective action.
Its 2-dim sections will be discussed at length in the following.
\bigskip
{\sect 3. Perturbations on the throat}
\smallskip
Near the extremal limit, the two essential features of the geometry are the
\af region and the attached throat. It is therefore crucial to study the
propagation of the fields along the throat. From the \ads form of the metric
can be expected that the fields are in some way confined into the throat,
due to the infinitely increasing gravitational potential for $\sigma\to
\infty$.

Given a perturbation $\chi$ on the throat, one can easily check that the
kinetic term in the action at the linearized level is
$$S_\chi=-\int\sqrt{\bar g}\ d^4x\ e^{-2b\phi}(\nabla\chi)^2,\eqno(3.1)$$
where $\bar g$ is the flat metric and $b$ is a constant depending on the
mode considered.

The effect of the \ld background can be seen  by defining the new
field $\ct=e^{-b\phi}\chi$. The linearized action becomes therefore, modulo
a total derivative,
$$S_\chi=-\int\sqrt{\bar g}\ d^4x\left[(\nabla\ct)^2+\ct^2(b^2(\nabla\phi)^2-b
\nabla^2\phi)\right]=$$
$$=-\int\sqrt{\bar g}\ d^4x\left[(\nabla\ct)^2+M^2(\phi)\ct^2\right],
\eqno(3.2)$$ where
$$M^2(\phi)= const\times\exp\left(-4{1+k\over 1-k}\phi\right),$$
is a space dependent mass term for $\ct$, which becomes infinite in the
region of weak coupling $e^{2\phi}\to 0$, where the excitations are thus
suppressed by an infinite mass gap. They are therefore not allowed to escape
to infinity, which is in our case the \af region.

Similar results have been obtained in [4] for the "canonical" metric
$\tilde g=e^{-2\phi}g$. In particular, these results can help to interpret the
vanishing of the temperature (2.5) in the extremal limit : a potential
barrier hinders the interaction between the black hole and the external fields,
creating a mass gap. In our treatment, the geometrical origin of
the barrier is however more evident.
\bigskip
{\sect 4. The two-dimensional effective theory}

{\ssect a) Dimensional reduction}
\smallskip
In order to investigate the quantum properties of the extremal \bh, it is
useful to study the 2-dim effective model obtained by retaining only the
radial modes of the 4-dim theory. This approximation is justified by the
fact that the \bg solution, as we have seen, reduces to the \pd of two 2-dim
metrics near the horizon. The 2-dim theory is renormalizable and is a
generalization of the model considered in [9].

We then start with a discussion of the classical aspects of the 2-dim effective
theory.\footnote{$^\dagger$}{These properties will be discussed in more detail
in a forthcoming paper [15]. Related 2-dim models are discussed also in
[16, 11].}
The action (2.1) can be dimensionally reduced by taking the angular
coordinates to span a 2-sphere of constant radius $Q$: the resulting action
is
$$S={1\over 2\pi}\int\sqrt g \ d^2x\ e^{-2\phi}\left[R+4(\nabla\phi)^2-{2\over
3}(\nabla\psi)^2+\lo-\lt\ \efp\right],\eqno(4.1)$$
where
$$\lo={2(q^2+3)\over 2q^2+3}{1\over Q^2}={3+k\over 2Q^2},\qquad
\qquad\lt={2q^2\over 2q^2+3}{1\over Q^2}={1-k\over 2Q^2}.$$
The ensuing equations of motion are:
$$2\nabla_\mu\nabla_\nu\phi-{2\over 3}\nabla_\mu\psi\nabla_\nu\psi=
\gmn\left[2\nabla^2\phi-2(\nabla\phi)^2-{1\over 3}(\nabla\psi)^2+{\lo\over 2}
-{\lt\over 2}\efp\right],$$
$$4\nabla^2\phi-4(\nabla\phi)^2-{2\over 3}(\nabla\psi)^2+\lo+R=0,\eqno(4.2)$$
$$\nabla^2\psi-2(\nabla\phi)(\nabla\psi)+{q\over 2}\lt\efp=0.$$

With the ansatz
$$e^{{2q\over3}\psi}={ q^2\over3}e^{2\phi},\eqno(4.3)$$
the 2-dim action admits the exact solution:
$$\ds-\sinh^2(\kappa\sigma)\cosh^{2k}(\kappa\sigma)dt^2+d\sigma^2,$$
$$\ef=\cosh^{k-1}(\kappa\sigma),\eqno(4.4)$$
where $\kappa={\lambda_2\over\sqrt{2(1-k)}}$, with $k$ defined by (2.3).

These solutions are everywhere regular for any value of $k$, and have a
horizon at $\sigma=0$. They of course coincide with the 2-dim section of the
solutions (2.10). For $k\ne-1$ the solutions (4.4) behave as \ads for
$\sigma\to\infty$. For $k=-1$ they are \af and reduce to those obtained in
[17].

Another class of solutions to (4.2,3) is given by the \ads \bg with \ld:
$$\ds-e^{2(k+1)\kappa\sigma}dt^2+d\sigma^2,$$
$$\phi -\phi_0=\kl\kappa\sigma,\eqno(4.5)$$
which we shall refer as ADS \ld vacuum. These solutions are asymptotic of
(4.4) for $\sigma\to\infty$ and correspond to the dimensional reduction of
(2.9).

We also notice that, substituting the ansatz (4.3) directly into the action,
one has:
$$S={1\over2\pi}\int\sqrt g\ d^2x\ e^{-2\phi}\left[R+{8k\over
k-1}(\nabla\phi)^2
-\lt\right].\eqno(4.6)$$
For $k=0$ the action reduces to that of the Jackiw-Teitelboim theory [18].
\bigskip
{\ssect b) The \Sch gauge}
\smallskip
In the following, it will be useful to write the metric in \Sch coordinates.
In such \coo it is in fact possible to continue the metrics besides the
horizon at $\sigma=0$ and to get a more immediate insight of their physical
properties.
The new \coo are defined so that the metric takes the form
$\ds -\Upsilon (r)dt^2+\Upsilon^{-1}(r) dr^2$.
The general solutions in this gauge are:
$$\ds(b^2r^2-a^2r^\kk)dt^2+(b^2r^2-a^2r^\kk)^{-1}dr^2,$$
$$\ef=(br)^{\ka},\eqno(4.7)$$
where $b={(1+k)\lambda_2\over\sqrt{2(1-k)}}$ and $a$ is a free parameter which,
from a 4-dim point of view, can be interpreted as parametrizing the departure
of the solution from extremality. These \coo can be related to the previous
ones by the transformation $br=\cosh^{1+k}(\kappa\sigma)$, with $a=b$.

The metric asymptotes \ads\st for $r\to\infty$, and has a horizon at $r_0=
\left({a\over b}\right)^{1+k}$, which shields a singularity at $r=0$, except in
the special cases $k=0, 1$. This is easily seen by considering the
curvature:
$$R=-2\left(b^2+{k(1-k)\over(1+k)^2}a^2r^\kd\right).\eqno(4.8)$$
If $a=0$, the metric reduces to that of 2-dim \ads\st with curvature $-2b^2$:
$$\ds-(br)^2dt^2+(br)^{-2}dr^2,\qquad\qquad\ef=(br)^{\ka},\eqno(4.9)$$
which corresponds to the solution (4.5).

The mass of the solutions can easily be obtained by the ADM procedure and is
given by:
$$M={1-k\over 2(1+k)}\ e^{-2\phi_0}a^2b^\kf,\eqno(4.10a)$$
or, in terms of the value $\phi_H$ of $\phi$ at the horizon,
$$M={\la\over 2}\sqrt\km\exp\left(2\kg\phi_0-{4\over 1-k}\phi_H\right)
.\eqno(4.10b)$$

Analogously, using standard procedures, one can calculate the thermodynamical
parameters:
$$T={1\over2\pi}e^{(1+k)\phi_0}\left({2M\over 1-k}\right)^\kp\left({b\over
1+k}\right)^\km,\eqno(4.11)$$
$$S=2\pi e^{(1+k)\phi_0}\left(\kg\ {2M\over b}\right)^\km,\eqno(4.12)$$
where the entropy $S$ is computed with respect to the asymptotic \ads
solution. The temperature increases monotonically with the mass and goes to
zero for $M=0$, which should therefore be considered as a stable \gs for the
quantum \bh. The entropy displays essentially the same behaviour as $T$. Also
is remarkable the relation $ST={2M\over 1-k}$, valid for $\phi_0=0$.

{}From the previous discussion is evident that for $a\ne 0$ all the metrics
(4.7) describe black holes with asymptotically \ads behaviour. There are
however
three special cases, which should be considered separately:
\bigskip
{\noindent 1) $k=-1$:}
$$\ds-(1-a^2e^{-br})dt^2+(1-a^2e^{-br})^{-1}dr^2,\qquad\qquad
\ef=e^{br}.\eqno(4.13)$$

This is the \af solution thoroughly discussed in [17]. Therefore we do not
discuss it here.
\bigskip
{\noindent 2) $k=0$:}
$$\ds-(b^2r^2-a^2)dt^2+(b^2r^2-a^2)^{-1}dr^2,\qquad\qquad\ef=(br)^{-1}
.\eqno(4.14)$$

For $a\ne 0$ the metric, while having a constant negative curvature has
different properties from \ads\st. In particular, it possesses a horizon at
$r=a/b$, but no singularities. We consider it in more detail below.
\bigskip
{\noindent 3) $k=1$:}
$$\ds-(b^2r^2-a^2r)dt^2+(b^2r^2-a^2r)^{-1}dr^2,\qquad\phi= const.\eqno (4.15)$$

In this case, the metric can be put in the same form as for the $k=0$ case,
by simply shifting $r$. Now, however, the dilaton is everywhere constant.
\bigskip
{\ssect c) 2-dim spacetimes of constant negative curvature.}
\smallskip
In order to discuss the properties of the $k=0$ solution and for future
reference, it is useful to summarize the properties of the 2-dim spacetimes of
constant negative curvature. They can be easily constructed by considering
\ph embedded in 3-dim \Mi\st, with metric $\ds dz^2-dx^2-dy^2$. The standard
\ads \st is then represented by a 1-sheet hyperboloid of equation [19]:
\footnote{$^\dagger$}{For simplicity we put $\Lambda =1$ in the following.}
$$x^2+y^2-z^2=1.\eqno(4.16)$$
The surface can be parameterized by the \coo:
$$x=\cosh\sigma \sin t,\qquad\qquad y=\cosh\sigma\cos t,\qquad\qquad
z=\sinh\sigma,\eqno(4.17)$$
with $-\infty<\sigma<\infty,\ 0\le t\le 2\pi$, giving rise to the metric
$$\ds-\cosh^2\sigma dt^2+d\sigma^2,\eqno(4.18a)$$
or, in the \Sch gauge,
$$\ds-(r^2+1)dt^2+(r^2+1)^{-1}dr^2.\eqno(4.18b)$$

Another parametrization, which however does not cover the whole hyperboloid,
is given by:
$$x=e^\sigma t,\qquad\qquad y=\cosh\sigma-{1\over 2}e^\sigma t^2,\qquad\qquad
z=\sinh\sigma+{1\over 2}e^\sigma t^2,\eqno(4.19)$$
with $-\infty<\sigma<\infty,\ -\infty<t<\infty$. In these \coo,
$$\ds-e^{2\sigma}dt^2+d\sigma^2,\eqno(4.20a)$$
or
$$\ds-r^2dt^2+r^{-2}dr^2.\eqno(4.20b)$$

Contrary to the 4-dim case, in 2 dimensions there is another \st of constant
negative curvature. This is represented by a 2-sheet hyperboloid of equation
$$x^2+y^2-z^2=-1,\eqno(4.21)$$
embedded as before in 3-dim flat \st.
With the choice of \coo:
$$x=\sinh\sigma \sin t,\qquad\qquad y=\sinh\sigma\cos t,\qquad\qquad
z=\cosh\sigma,\eqno(4.22)$$
$0\le\sigma<\infty,\ 0\le t\le 2\pi$, the metric becomes\footnote{$^\dagger
$}{In this form, the metric has been independently discussed in [20],
in the context of the Jackiw-Teitelboim theory [18]}:
$$\ds-\sinh^2\sigma dt^2+d\sigma^2,\eqno(4.23a)$$
or, in \Sch\coo ,
$$\ds-(r^2-1)dt^2+(r^2-1)^{-1}dr^2.\eqno(4.23b)$$

To our knowledge, this metric has not been previously discussed in the
literature, presumably because it cannot be obtained by a \dr of 4-dim
\ads\st (see however [20]).

{}From its construction, it is evident that the solution is everywhere regular.
Its most striking difference from \ads\st is the presence of a horizon at
$r=1$ (the vertex of the hyperboloid). The asymptotic properties instead, are
identical in both cases. Thus the metric describes a non-singular \bh. Its
horizon
has temperature $T={1\over 2\pi}$ at variance with the \ads metric (4.18),
whose temperature of course vanishes.
We shall discuss in extent this feature in the following sections.

To conclude we recall some problems in the interpretation of \ads-like metrics
which emerge when a quantum theory is constructed on such \bg [22]. First
of all, from the geometry of the \st and the definitions (4.17), (4.22) is
evident that the time coordinate is periodic, $0\le t\le 2\pi$.  The consequent
presence of closed timelike paths can however be avoided by going to the
universal covering space, $-\infty<t<\infty$.
The second problem is the lack of global hyperbolicity of \ads\st, due to the
fact that spacelike infinity is at finite coordinate distance in conformal
coordinates. A great care is therefore needed in the choice of the \bc [22].
We shall discuss at length these problems in section 6.

\bigskip
{\sect 5. The solutions in the conformal gauge}

{\ssect a) The general solution}
\smallskip
It is useful to write the two-dimensional \bh solutions of the previous
section in the conformal gauge.
In this  gauge
 $$ds^2=-e^{2\rho}d\xn d\xm,\qquad x^\pm=x^0\pm x^1,  $$
the field equations (4.2) become
$$\eqalign{\dd\dm\rho&=2 \dd\dm\phi-2 \dd\phi\dm\phi-{1\over3}
\dd\psi\dm\psi - {\lambda^2_1\over 8}e^{2\rho},\cr
\dd\dm\phi&=2 \dd\phi\dm\phi + {\lambda^2_1\over 8}e^{2\rho} -{\lambda^2_2
\over 8}
e^{2(\phi +\rho -{q\over3}\psi)},\cr \dd\dm\psi&=\dd\psi\dm\phi+\dd\phi\dm\psi
+q{\lambda^2_2\over 8}e^{2(\phi +\rho -{q\over3}\psi)},\cr}\eqno(5.1)$$
to be solved under the constraints
$$\eqalign{\dd^2\phi-2 \dd\rho\dd\phi-{1\over 3}(\dd\psi)^2&=0,\cr
\dm^2\phi-2 \dm\rho\dm\phi-{1\over 3}(\dm\psi)^2&=0.\cr}\eqno(5.2)$$
The ansatz (4.3) reduces the previous equations to the form
$$\eqalign{\dd\dm\rho&={1+3k\over k-1} \dd\dm\phi+2\kr \dd\phi\dm\phi,\cr
\dd\dm\phi&=2 \dd\phi\dm\phi + {\lambda^2_2\over 8}e^{2\rho},\cr}\eqno(5.3)$$
$$\eqalign{\dd^2\phi-2 \dd\rho\dd\phi-2\kr(\dd\phi)^2&=0,\cr
\dm^2\phi-2 \dm\rho\dm\phi-2\kr(\dm\phi)^2&=0,\cr}\eqno(5.4)$$
where we have introduced the parameter $k$ given by eq. (2.3).
The constraints (5.4) may be solved for $\rho$ in terms of $\phi$
$$\rho={1\over 4} \ln\left(-{1\over \lambda^2_2}\dd\phi\dm\phi\right)-
\kr \phi +w_++w_-,$$
$$\dd\phi=-e^{4(w_+-w_-)}\dm\phi,$$
where $w_{+}(\xn), w_-(\xm)$ are two arbitrary functions of the two light-cone
variables.
Fixing the residual gauge freedom relative to the conformal subgroup
of diffeomorphisms by setting $w_{\pm}=0$, one can integrate the equation
of motion (5.3). The general solutions describing black holes
with a regular horizon are
$$\eqalign{e^{2\rho}={1\over \la }&\left[A \exp(4\kr X)- C \exp(4\ks X)
\right],\cr \ef &=e^{-2X},\cr}\eqno(5.5)$$
where $X=X\ma$ is defined implicitly by
$$\ma=-\int_{\infty}^{X}dX'\left[A\exp(2\kr X')-C \exp(-2X')\right]^{-1},
\eqno(5.6)$$
$$A={\la\over 32}(1-k),\eqno(5.7)$$
and $C\ge0$ is an integration constant.
Eqs. (5.5) represent the \bh solutions (4.7) written in the conformal gauge.
The singularity is located at $X\rightarrow -\infty$, the horizon at
$X={1-k\over 4}
\ln\left(C\over A\right)$. The region of weak couplings ($e^\phi =0$)
corresponds to $X\rightarrow\infty$.
The constant $C$ appearing in (5.5,6) is related to the \bh mass.
In fact using the formula (4.10b) we get for the ADM mass of the hole
$$M=8\sqrt{2\over 1-k}e^{-2\phi_0} C.\eqno(5.8)$$
The zero-mass solution ($C=0$) corresponds to the ADS linear dilaton
vacuum (4.5).
In this case we can perform explicitly the integral (5.6) and write
the solution (5.5) in the form
$$e^{2\rho}={8\over\la^2}{1-k\over(1+k)^2}\ma^{-2},\eqno(5.9)$$
$$\ef=\left[\la{k+1\over16}\ma\right]^{1-k\over1+k}.\eqno(5.10)$$
The Kruskal diagram for  this solution is shown in fig. 1.
Note that although (5.9) is defined over the whole range of the
coordinates, (5.10) exists only for $\xm\ge\xn$, i.e. in the left
hand side of fig. 1. The line $\xm=\xn$ corresponds to the region
of weak couplings $e^\phi=0$, i.e. to the asymptotical $\sigma\rightarrow
\infty$ region of the solution (4.5).
The solutions defined for $\xn\ge\xm$ can be obtained by changing the signs
of $\xn$ and $\xm$.  In the following we will only consider solutions
defined for $\xm\ge\xn$.
The isometry group of the metric (5.9) is $GL(2,R)$ realized as the
fractional transformations
$$\xn\rightarrow {a\xn+b\over c\xn+d},\qquad\xm\rightarrow {a\xm+b\over c\xm+d}
,\eqno(5.11)$$
with $ad-bc\not=0$. The dilaton solution (5.10) however is not invariant
under this transformations.
 \bigskip
{\noindent\sl b) The k=0 case}
\smallskip
In section 4.a) we have seen how for $k=0$  our model becomes the theory
of Teitelboim and Jackiw [18]. It is also interesting to write the
corresponding
solutions in the conformal gauge.
For $k=0$ the equation of motion (5.3) can be written as
$$\eqalign{\dd\dm\rho&=-{\la^2\over8}e^{2\rho},\cr
\dd\dm\phi&=2 \dd\phi\dm\phi+ {\la^2\over8}e^{2\rho},\cr}\eqno(5.12)$$
whose general solution appears in the literature in the form [21]:
$$\eqalign{e^{2\rho}&=\left(1+{\la^2\over32}\xn\xm\right)^{-2},\cr
e^{-2\phi}&={\alpha_0\left(1-{\la^2\over32}\xn\xm\right)+\alpha_+\xn+\alpha_-
\xm\over 1+{\la^2\over32}\xn\xm},\cr}\eqno(5.13)$$
where $\alpha_0, \alpha_\pm$ are arbitrary constants.
The solutions (5.13) can be written,  using the residual coordinate invariance
within the conformal gauge, in a form similar to the ADS linear dilaton vacuum
(5.9,10)
 $$e^{2\rho}={8\over\la^2} \ma^{-2},\eqno(5.14)$$
$$\ef=\la{\ma\over 1-\alpha^2\xn\xm},\eqno(5.15)$$
where $\alpha$ is a free parameter related, as we shall see in the following,
to the \bh mass.
The solutions (5.14,15) describe a non singular \bh. In fact the curvature
tensor
(4.8) is everywhere regular whereas an horizon appears where ${(\nabla\phi)}^2$
change sign, i.e. for
$$\xn=\pm{1\over\alpha}, \qquad \xm=\pm{1\over\alpha}.$$
The maximally extended spacetime has the structure of a chain of connected
multiuniverses [20].
The ADM mass of the hole is
$$M={\sqrt2\over4}e^{-2\phi_0}{\alpha^2\over\la}.\eqno(5.16)$$
For $\alpha=0$ (zero-mass hole) we find the ADS linear dilaton vacuum
(5.9,10).
Notice that the solutions describing a \bh with a given mass $M$ differ
from the zero-mass solution just for the value of the dilaton, the metric
part being unchanged.
The Kruskal diagram of these solutions is shown in figure 2.
The bold line, characterized by $\xn=1/\alpha^2\xm$ does not represent
a singularity of the metric but the line where the theory becomes strongly
coupled ($e^\phi\rightarrow\infty$).
A zero-mass particle, conformally coupled to two-dimensional gravity,
coming from the region of weak couplings ($\xn=\xm$) will hit this line
and without encountering a singularity will jump in another universe.
It is worth noting that the solutions (5.14,15) are invariant under the
duality transformations
$$\xn\rightarrow{1\over\alpha^2\xn}, \qquad
\xm\rightarrow{1\over\alpha^2\xm},$$
which are a subgroup of the full isometry group $GL(2,R)$ of the metric.
\bigskip
{\sect 6.  Hawking radiation}

{\ssect a) Schwarzschild gauge}
\smallskip
We have seen in the previous sections that our classical solutions represent
two-dimensional black holes in anti-de Sitter (ADS) space-time. At the quantum
level one expects these black holes to evaporate in the same way as black holes
in flat spacetime do.
However in our case the discussion of the Hawking effect is complicated
by the subtleties of the quantization of fields propagating in a ADS background
[22].
Anti-de Sitter spacetime is not globally hyperbolic, information can be lost to
or gained from
spatial infinity in finite coordinate time. The effect of this loss or gain of
information on the Cauchy problem is even worse in the case at hand because
we do not have full control of the boundary conditions at spatial infinity.
In fact our two-dimensional solutions describing an ADS throat, have to be
reembedded in a four-dimensional,  asymptotically flat, manifold. The details
of this reembedding are crucial for setting the appropriate boundary conditions
for the two-dimensional problem. We know, however,  that a consistent
quantization of scalar fields in anti-de Sitter space-time exists [22].
Both "transparent" or "reflective" boundary conditions may be used,  the
information being either "recycled" or reflected at spatial infinity.
When dealing with a complete space-time such as ADS no other possibility is
left. Our infinite throat,  however, is embedded in a four-dimensional
space-time.  One way of establishing a well defined Cauchy problem is to
accept that our two-dimensional manifold is incomplete and require
the Cauchy data to be specified on a Cauchy surface of the surrounding
spacetime.  We do not know exactly how to do this because we are not able to
explicitly describe the reembedding of the two-dimensional space in the four
dimensional one. Nevertheless, in principle, information should be allowed
to leave the two-dimensional anti-de Sitter space-time and to reach the
asymptotically flat region of the four-dimensional space. In conclusion
we are left with two main sets of boundary conditions for our
two-dimensional problem: the information is either allowed to leave the
two-dimensional spacetime or is reflected at spatial infinity (we will not
consider here the rather exotic possibility that the information is
"recycled" at spatial infinity).

As a first step in the study of the Hawking effect let us discuss the
Hawking temperature associated with the horizon of the solutions (4.7).
As we have seen in section 4 b) the temperature of the hole, eq. (4.11),
goes to zero with the mass.
This indicates that,  loosing mass through the Hawking radiation,  the
\bh will set down to a zero-mass non radiating, stable ground state which
could be naturally interpreted as the ADS linear dilaton vacuum.
This is strictly true only if one uses boundary conditions which allow
the radiation to escape at spatial infinity. Reflecting boundary conditions
will make also possible a ground state describing a \bh in thermal equilibrium
with the radiation. The local measured temperature is red-shifted
by the ADS gravitational potential and decreases the further one is from
the horizon.

One way for extracting more information about the Hawking evaporation
process is to use,  following the lines of ref. [23],  the covariant
conservation equations
$$\nabla_\nu T_\mu^\nu=0,\eqno(6.1)$$
for the stress tensor $T_\mu^\nu$ in the background defined by (4.7).
In the two-dimensional, asymptotically flat case this gives the well known
relation between the Hawking black body effect and the trace anomaly for
2D-conformal matter coupled to gravity [23].
We shall integrate the conservation equations (6.1) in the conformally flat
background
$$ds^2=\Upsilon\left(-dt^2+dr_*^2\right),\eqno(6.2)$$
which is the metric (4.7) expressed in terms of the coordinate $r_*$,
defined by
$${dr\over dr_*}=\Upsilon,$$
with
$$\Upsilon= b^2r^2-a^2r^{2k\over k+1}.$$
The resulting form of $T_\mu^\nu$ is the following
$$T_\mu^\nu=T_\mu^{(1)\nu}+T_\mu^{(2)\nu}+T_\mu^{(3)\nu},\eqno(6.3)$$
with
$$T_\mu^{(1)\nu}=\left(\matrix{-\Upsilon^{-1}H_2+T_\alpha^\alpha&0\cr0&
\Upsilon^{-1}H_2\cr}\right),\eqno(6.4)$$
$$T_\mu^{(2)\nu}={K\over\Upsilon}\left(\matrix{1&-1\cr1&-1\cr}\right),
\eqno(6.5)$$
$$T_\mu^{(3)\nu}={J\over\Upsilon}\left(\matrix{-1&0\cr0&1\cr}\right),
\eqno(6.6)$$
 where
$$H_2={1\over2}\int_{r_o}^r dr' T_\alpha^\alpha(r')\partial_{r'}\Upsilon,$$
$r_0$ is the location of the horizon and $J,K $ are integration constants to be
determined by boundary conditions.
Regularity of the stress tensor on the future horizon requires $J=0$.
To be more concrete let us consider $N$ massless matter scalar fields
conformally coupled to 2D-gravity. When the scalar fields are quantized
in the classical background geometry (6.2) a conformal anomaly will
show up.
The conformal anomaly results in the non-vanishing trace for the stress
tensor [24]:
$$T_\alpha^\alpha={N\over96}R.\eqno(6.7)$$
Using the expression (4.8) for the curvature tensor $R$ one can easily
calculate the asymptotic form ($r\rightarrow\infty$) of the stress
tensor (6.3)
$$\left(T_\mu^\nu\right)_{as}=-{N\over 192}{(k+1)^2\over 1-k}\la^2\delta_
\mu^\nu+T_\mu^{(2)\nu},\eqno(6.8)$$
with $T_\mu^{(2)\nu}$ given by eq. (6.5).
The first term in eq. (6.8) represents the quantum correction to the
vacuum energy of the anti-de Sitter background,  whereas the second term
describes the flux of Hawking radiation. The flux is red-shifted
by the factor $\Upsilon^{-1}$ and is actually zero at infinity.  For this
reason one cannot use the boundary conditions at spatial infinity in order to
determine the constant $K$ as one usually does for asymptotically flat
spacetime.  The expression (6.8) gives a time-independent description of
the Hawking flux.  It describes the Hawking radiation long after the matter
has collapsed to form a black hole. In this context nothing can be said about
the end point of the \bh evaporation.
More details about the Hawking evaporation process can be achieved by
considering the dynamics of a collapsing massless shell.  This will
be the subject of the following section.
\bigbreak
\bigskip
{\ssect b) Collapse of a massless shell}
\smallskip
Consider $N$ scalar massless fields $f_i$ conformally coupled to
the 2D-gravity model defined by the action (4.1).
The classical action is
$$S={1\over2\pi}\int d^2x\sqrt g\left\{\left[R+4(\nabla\phi)^2-{2\over 3}
(\nabla\psi)^2+\lambda_1^2-\la^2 e^{2(\phi-{q\over3}\psi)}\right]e^{-2\phi}
-{1\over2}\sum_{i=1}^N(\nabla f_i)^2\right\}.\eqno(6.9)$$
Since we are dealing with massless fields  we will use the
conformal gauge to study the corresponding dynamics.
The classical equation of motion for the metric, the dilaton and the
field $\psi$ are the same as for the matter-free case. They are given
by eqs. (5.1). The field equations for the matter fields and the constraints
are
$$\dd\dm f_i=0,\eqno(6.10)$$
$$\eqalign{e^{-2\phi}\left[\dd^2\phi-2 \dd\rho\dd\phi-{1\over 3}(\dd\psi)^2
\right]&={1\over 4}\sum_{i=1}^N\dd f_i\dd f_i,\cr
e^{-2\phi}\left[\dm^2\phi-2 \dm\rho\dm\phi-{1\over 3}(\dm\psi)^2
\right]&={1\over 4}\sum_{i=1}^N\dm f_i\dm f_i.\cr}\eqno(6.11)$$
The gravity system defined by the action (4.1) possesses one local degree
of freedom. The question of stability becomes quite non trivial
because there are several classical time-dependent matter-free solutions.
However we look for solutions verifying the relation (4.3).
Once the field $\psi$ as been "frozen" by means of eq. (4.3) we are left
with a system devoid of local degrees of freedom. The solutions of the
matter-free equations of motion (5.3) are therefore locally static and for
a shell of collapsing matter  we can find the
solutions in a way similar to that of ref. [9], i.e. by gluing together
different solutions of the matter-free system across the history of the
shell.
Moreover the two-dimensional action (4.1) comes from a compactification
on a four-dimensional background which does verify  the relation (4.3).
One is thus forced to consider also for the two-dimensional gravity
system classical orbits for which eq. (4.3) holds.

Following ref. [9] we will consider solutions describing an $f$-shock wave
at $\xn =\xn_0$ travelling in the $\xm$ direction, whose only non-vanishing
stress tensor component is
$$T_{++}= {1\over2}\sum_{i=1}^N\dd f_i\dd f_i=2\beta\delta(\xn-\xn_0),\eqno
(6.12)$$
where $\beta$ is the magnitude of the wave.
Substituting eq. (6.12) in eq. (6.11) one easily finds  the solutions
of  the equation of motion (5.3) and of the constraints (6.11) which are
continuous across the shock $\xn=\xn _0$.
They are given, for $\xn\le\xn_0$,  by the ADS linear dilaton vacuum
$$\eqalign{e^{2\rho}&={8\over\la^2}{1-k\over(1+k)^2} \ma^{-2},\cr
\ef&={\left[{\la (k+1)\over 16}\ma\right]}^{1-k\over 1+k}
,\cr}\eqno(6.13)$$
and for $\xn\ge\xn_0$ by ($'=d/d\xm$)
$$\eqalign{e^{2\rho}=F'(\xm)&{1\over \la}\left[A \exp(4\kr X)- e^{2\phi_0}
\beta \exp(4\ks X)\right],\cr
\ef &= e^{-2X},\cr}\eqno(6.14)$$
where $X=X(x^+,x^-)$ is defined implicitly by
$$\xn-\xn_0-F(\xm)=\int_{\infty}^{X}dX'\left[A\exp(2\kr X')-e^{2\phi_0}\beta
\exp(-2X')\right]^{-1},\eqno(6.15)$$
$$F(\xm)=\int _{\xn _0}^{\xm} dy^-\left[1-\gamma\left(y^- -\xn _0
\right)^{2\over1+k}\right]^{-1},\eqno(6.16)$$
$$\gamma=e^{2\phi_0} \beta\left[2(1+k)\over1-k\right]^{2\over1+k}A^
\kq.\eqno(6.17)$$
The constant $A$ is given by eq. (5.7). The function $F(\xm)$ has been
chosen in the form (6.16) in order to fulfil the required continuity
conditions for $\rho$ and $\phi$ at $\xn=\xn_0$.
Comparing eqs. (6.14-17) with eqs. (5.5-7) one easily realizes that the
redefinition $\xm\rightarrow F^{-1}(\xm)$ brings the solutions
(6.14) in the form (5.5). Thus eqs. (6.14) describe a black hole of mass
$M\propto\beta$. As expected the  effect of the matter perturbation  (6.12)
on the ADS linear dilaton vacuum (6.13) is to produce a \bh of mass
proportional to $\beta$.
\bigskip
{\ssect c) Hawking radiation in the conformal gauge}
\smallskip
In  section 6.b) the discussion has been purely classical.
The Hawking radiation is a quantum effect which will appear when
the matter fields are quantized in the classical background geometry defined by
eqs (6.13,14).  In the quantum theory the matter fields $f$ couple to
the gravitational degrees of freedom owing to the conformal anomaly.
At the one-loop level,  ignoring the back-reaction of the metric,
the effect of the conformal anomaly can be summarized in the non
vanishing trace of the stress tensor [9]:
$$< T_{+-}>=-{N\over12}\dd\dm\rho.\eqno(6.18)$$
Integrating the equation of conservation for $T$ one obtains the following
one loop expressions for $< T_{++}>$ and $< T_{--}>$ [9]:
$$\eqalign{< T_{++}>&=-{N\over12}\left[(\dd\rho)^2-\dd^2\rho+t_+(\xn)
\right],\cr<T_{--}>&=-{N\over12}\left[(\dm\rho)^2-\dm^2\rho+t_-(\xm)
\right].\cr}\eqno(6.19)$$
The functions $t_{\pm}$ appearing in eqs. (6.19) have to be fixed
by means of the boundary conditions.
Using eqs. (6.13-17) one obtains
$$\eqalign{< T_{++}>&=-{N\over12}t_+(\xn),
\cr<T_{--}>&=-{N\over12}t_-(\xm),
\cr {\rm for}\quad &\xn \le\xn_0,\cr}\eqno(6.20)$$
and
$$\eqalign{< T_{++}>&={N\over12}\left[P+{2k\over k-1}e^{2\phi_0}\beta
e^{-2X}\delta(\xn-\xn_0)-t_+\right],\cr< T_{--}>&={N\over12}\left[
(F')^2\ P+{1\over2}\left\{F,\xm\right\}0-t_-\right],\cr
{\rm for}\quad & \xn\ge\xn _0,}\eqno(6.21)$$
where
$$P={4\over k-1}e^{2\phi_0}\beta A\exp({4k\over1-k}X)-{4k\over{(k-1)}^2}
e^{4\phi_0}
\beta^2\exp(-4X),\eqno(6.22)$$
and $$\left\{F,\xm\right\}={F'''\over F'}-{3\over2}{\left({F''\over F'}
,\right)}^2\eqno(6.23)$$
is the Schwarzian derivative of the function $F$.
We will fix the boundary conditions by requiring that $T$ vanishes identically
in the ADS linear dilaton region, which in turn implies $t_-=0$, and that
there should be no incoming radiation at spatial infinity for every value of
$\xm$ except for the classical $f$-wave at $\xn=\xn_0$.
Let us discuss the physical meaning of the expression (6.21) for the different
values of the parameter $k$.
\medskip
{\ssect 1. $k=0$}
\smallskip
One easily finds
$$< T_{--}>=0 ,\qquad {\rm identically}.$$
The fact that $<T_{--}>$ vanishes identically over the whole space
means that there is no Hawking radiation at all.  This result is not surprising
because,  as we have seen in section 5.b), the solutions (5.14-15) describe
non singular black holes. Moreover the metric part of the solutions
(5.14) is the same as for the ADS linear dilaton vacuum. Thus at the
classical level the effect of the shock wave on the ADS linear dilaton
background is encoded in the modification of the dilaton given by eq. (5.15).
The absence of Hawking radiation is the consequence of the insensitivity of
the metric part of the background to the shock wave.
\medskip
{\ssect 2. $1< k< 0$}
\smallskip
The flux of Hawking radiation is given by the asymptotical value of
$<T_{--}>$ as  $X\rightarrow\infty$.
Taking into account that in  (6.21) $P\rightarrow0$ as  $X\rightarrow\infty$
one obtains the expression
$$< T_{--}>_{as}={N\over24}\left\{F,\xm\right\}.\eqno
(6.24)$$
Using the expression (6.16) for $F$ eq. (6.24) becomes
$$< T_{--}>_{as}={N\gamma\over 12{(1+k)}^2}
{\left[(1-k)(\xm-\xn_0)^{-2k\over k+1}+\gamma k(\xm-\xn_0)^{2(1-k)\over k+1}
\right]\over{\left[1- \gamma (\xm-\xn_0)^{2\over1+k}\right]^{2}}}.
  \eqno(6.25)$$
Let us discuss the expression (6.25).
First we note that $< T_{--}>_{as}$ blows up at the
horizon,  i.e. for
$$\xm=\xm_1=\xn_0+ \gamma^{-(1+k)\over2}.\eqno(6.26)$$
This divergence does not have an invariant physical meaning.
In fact it is a consequence of the bad behaviour of our coordinate system
on the horizon. The function $F$ of eq. (6.16) diverges for $\xm=\xm_1$.
This divergence can be easily eliminated by going back to the original
coordinate system of eq. (5.5,6), i.e by defining the new light-cone
coordinate
$$\tilde\xm=F(\xm).$$
Expressing the tensor $< T_{--}> $ in this new coordinate system we
have
$$< \tilde T_{--}> =(F')^{-2}< T_{--}> -{N\over12}\left[{1\over 2(F')^2}
\left\{ F,\xm\right\}+\tilde t_-(\xm)\right],\eqno(6.27)$$
where $<T_{--}>$ is given by eq. (6.20) and (6.21) (with $t_-=0$) for
the  regions $\xn\le\xn_0$ and $\xn\ge\xn_0$ respectively.
Requiring that $< \tilde T_{--}>$ vanishes in the ADS linear dilaton region
and taking into account that there $<T_{--}>=0$, one can identify the
function $\tilde t_-(\xm)$
 $$\tilde t_-(\xm)=-{1\over 2(F')^2}\left\{F,\xm\right\}.\eqno(6.28)$$
With this position  we can obtain, using eq. (6.24), the Hawking flux
$$<\tilde  T_{--}>_{as}={N\over24}{1\over (F')^2}\left
\{F,\xm\right\}.\eqno(6.29)$$
Finally substitution of $F$ gives the result
$$<\tilde T_{--}>_{as}={N\over12}{\gamma\over{(1+k)}^2}
\left[(1-k)(\xm-\xn_0)^{-2k\over k+1}+\gamma k(\xm-\xn_0)^{2(1-k)\over
k+1}\right] . \eqno(6.30)$$
The expression for the Hawking flux is now well behaved for every finite
value of $\xm$. The Hawking flux starts at the value
$$<\tilde  T_{--}>_{as}=0,\qquad {\rm for}\quad\xm=\xn_0,$$
and reaches  at the horizon $\xm=\xm_1$ its  maximum :

$$<\tilde T_{--}>_{as}={N\over12}{\gamma ^{1+k}\over{(1+k)}^2}.\eqno(6.31)$$

For $\xm\le \xn_0$, $<\tilde T_{--}>_{as}$ vanishes identically because
there we are in the ADS linear dilaton region.
Recall that in our coordinate system spatial
infinity is identified with the line $\xm=\xn$. We are therefore forced
to send the shock wave adiabatically at $\xm=\xn_0$.
Inserting the expression (6.17) for $\gamma$ in eq. (6.31) one easily
finds the following behaviour of the Hawking flux on the horizon
 $$<\tilde T_{--}>_{as}\propto\beta^{1+k}\la^{1-k}\propto M^{1+k}\la^{1-k},
\eqno(6.32)$$
$M$ being the \bh mass. Thus the value of the Hawking flux on the horizon goes
to zero with the mass of the hole. This behaviour has to be compared with
previous results for the Hawking flux in 2-dim dilaton gravity where an
asymptotically mass independent Hawking radiation rate has been found [9].
As stated in sect. 4 the 2-dim dilaton gravity theory of ref. [9] is the
particular case  $k=-1$ of our model. Inserting this value of $k$ in (6.32)
one finds
$$<\tilde T_{--}>_{as}\propto\la^2,$$
i.e the asymptotically mass independent Hawking radiation rate
found in ref. [9].

The fact that the Hawking flux at the horizon  goes to zero with the
mass of the hole strongly suggests that,  loosing mass through the
radiation, the \bh settles down to a stable, non radiating ground state.
This conclusion is of course the only possible if,  as we have seen
in sect. 6.a, the temperature of the hole goes to zero with the mass.
However, in the context of our semiclassical calculations,  which neglect
the back-reaction of the geometry, the previous result can not be seen
as a conclusive statement about the end point of the Hawking evaporation
process. In fact, when the black hole has radiated away most of its
initial mass, the back-reaction cannot be neglected.
A check of the range of validity of our approximation can be done by
calculating
the value of the weak field expansion parameter $e^{\phi}$ on the point
where the $f$-wave meets the horizon.
This point has evidently coordinates $(\xn_0,\xm_1)$ and a simple calculation
shows that
$$e^\phi\propto\left(\la\over\beta\right)^{1-k\over 4}\propto
\left(\la\over M\right)^{1-k\over 4}.$$
Thus for $\beta\gg\la$ or equivalently for $M\gg\la$, $e^\phi$ is a small
number. As expected for macroscopical holes ($M\gg\la$) the Hawking process
is seen to take place  in the region of weak couplings where one is
allowed to neglect the back-reaction of the metric.
For $M\sim\la$ our calculation of the Hawking flux breaks down and we have
to take into account the back-reaction in order to get meaningful
results. Unfortunately we are not able to solve the system of differential
equations describing the model when the back-reaction of the metric is taken
into account.
In conclusion, even though our calculations strongly indicate the emergence
of a stable ground  state at the end point of the evaporation process,
a final word on the subject deserves further investigation.
\medskip
{\ssect 3. $0< k< 1$}
\smallskip
For $k>0$ the function $P$ in eq. (6.21) diverges as $X\rightarrow\infty$.
Both $< T_{--}>$ and  $< T_{++}>$ will get therefore {\rm negative}
divergent asymptotical contributions. The meaning of these divergences will
be clarified shortly.

\smallskip
In section 4.b), we found a non-vanishing value for the temperature of the
hole for every value of the parameter $k$. Moreover, the temperature was seen
to go to zero with the mass. This seems in contradiction with the results
of this section where we have found a zero and a negative divergent
Hawking flux for $k=0$ and $k> 0$ respectively.
The solution of the puzzle lies in a careful analysis of the "quantum
vacuum" for fields propagating in an ADS background. It is well known
from the general theory of quantization in a curved background [25], that for a
given background geometry,  in general the choice of a particular
coordinatization of the space will result in a particular choice for
the quantum vacuum.  In particular the conformal vacua belong to classes
which have similar geometries but different topologies. Moreover, the classes
are related by thermalization at a given temperature in the sense that an
observer
sitting in a vacuum corresponding to a coordinatization will regard the vacuum
corresponding to the other coordinatization as a thermal bath at this
temperature.
Now the key point is that the vacuum defined by (4.14) is the thermalization
of the vacuum defined by  (5.14,15) at the temperature
$$T={1\over4\pi} \la^2 r_0.\eqno(6.33),$$
$r_0$ being the location of the horizon in  \Sch coordinates.
One can understand this by the following reasoning.
The metrics (4.9) and (4.14), when written in the conformal gauge, can  both
be put in the form (5.9). However the coordinate transformation required is
different. The region $r> 0$ for these two metrics is mapped by the
coordinate transformation in two different regions of the Kruskal diagram of
figure 1. Thus even though the metrics (4.9) and (4.14) have the same
analytical expression in terms of light-cone  coordinates,  they correspond to
different topologies. The relationship between the two corresponding
spacetimes  is analog to the one between  two-dimensional Rindler and \Mi
spacetime.
The \bh solution (4.14) has a natural temperature associated with it given
by eq. (6.33) while the ADS linear dilaton vacuum (4.9) has zero temperature.
The two corresponding quantum vacua are therefore related by thermalization at
temperature (6.33). Thus an observer sitting in the vacuum defined by the
metric (4.14) would regard the vacuum defined by the metric (5.14) as a
thermal state at this temperature. The temperature of the black hole (5.14)
is therefore zero.
The previous discussion explains not only why for $k=0$ in the conformal gauge
we have no Hawking radiation but also the meaning of the divergencies we found
for $k>0$. A simple consequence of the previous analysis is that in the
conformal gauge the black holes with $k=0$  have the same temperature
(equal to zero) as the ADS linear dilaton vacuum. Looking at formula (4.11)
one immediately
realizes that the \bh with $k> 0$ have temperature lower than (6.33).
They have, therefore, when described in the conformal gauge,
a temperature lower than the one associated with the ADS linear dilaton
vacuum i.e. a negative temperature. Thus  for $k> 0$  the ADS linear dilaton
vacuum is not the true ground state, it is essentially unstable and will tend
to
collapse to form a \bh.
\bigskip
{\sect 7.  Summary and outlook}
\smallskip
In this paper we have studied the extremal limit of the four dimensional \bh
solutions of the effective string theory which is obtained when in the
low-energy spectrum both the dilaton and a
modulus coming from the compactification of the string to 4-dim are present.
These 4-dim \bh solutions evidenciate  geometrical and thermodynamical
properties slightly different from the purely dilatonic case. In particular,
for extremal holes the temperature  approaches monotonically
to zero, indicating that the extremal limit has to be interpreted as
a non-radiating  ground state. One of the main results
of this work is that this behaviour has  a geometrical explanation in terms
of the underlying two-dimensional effective theory. In fact we found that near
the extremal point, the model is well described by an effective 2-dim gravity
theory whose \bh solutions are asymptotically anti-de Sitter rather then
asymptotically flat as for the purely dilatonic case. Thus the anti-de Sitter
gravitational potential acts as a confining one, being responsible for the
creation of a mass gap which hinders the interaction between
the \bh and the external fields.
Moreover the \bh solutions of the 2-dim gravity theory have a rather rich
structure. They can be considered as the generalization  of the
two-dimensional \bh solutions found in the purely dilatonic theory.
For the particular value $k=-1$ of the parameter characterizing the solutions
we find the asymptotically flat solutions of ref [17], whereas for $k=0$
our model becomes the Teitelboim-Jackiw theory whose solutions describe
non singular black holes. For generic values of $-1<k<1$ our solutions
describe 2-dim black holes in an anti-de Sitter background. This asymptotical
behaviour  of the solutions not only explains the peculiar thermodynamical
properties of the 4-dim black holes but has also important consequences
for the thermodynamics of the 2-dim black holes. In fact we found that
differently from the purely dilatonic case, the  temperature of the 2-dim
hole goes to zero with the mass indicating the emergence of a stable ground
state. This conclusion has been substantially confirmed by the study of the
Hawking evaporation process. In fact we have found, for a wide range of values
of the
parameter $k$, that the Hawking radiation rate is not asymptotically mass
independent as in the case of pure dilatonic 2-dim gravity but goes to zero
with the mass of the hole.
Even though in the calculations some subtleties are involved, mainly
connected with the difficulty in  setting the appropriate boundary
conditions and even though, having neglected the back-reaction of the geometry,
our study of the Hawking evaporation process is not complete  it seems to
us that our main results could hardly be spoiled by further investigations.

We conclude with some comments about the relevance of our results for the
problem of information loss in black hole physics.
One of the main motivations for the study of string inspired black hole
models is to find a solution to the information loss puzzle in the \bh
evaporation process. One possible solution is the idea that the information
is retained by a stable \bh remnant. This proposal in the context
of dilatonic 2-dim gravity  has been discussed at length in the literature
[7]. The information that resides in the \bh remnant is contained in the
infinite, narrow throat attached to the spacetime which characterizes the
magnetically charged \bh solutions of string theory.
However if the \bh remnants have to be the solution of the puzzle,
one has to solve two main problems.
First, the temperature and the rate of emission of the limiting
2-dim \bh  of purely dilatonic gravity does not go to zero. As pointed
out by Hawking in ref. [26] this is in contradiction with the fact
that the \bh settles down to a stable state. Of course one can make appeal to
the back-reaction of the geometry and/or to quantum gravity effects to halt
the evaporation process, so that a stable \bh remnant is left behind.
The 2-dim black holes we have discussed in this paper represent an improvement
in this direction. In fact they have a temperature and a rate of emission
which in the limiting case goes to zero. The idea that the \bh settles down
to a stable state looks more natural in this context.
Second, since the remnant must be able to encode the  information from an
arbitrarily large initial \bh, it must have an infinite spectrum. Owing to this
infinite spectrum all the ordinary physical processes would produce bursts
of remnants. It has been suggested that certain remnants in string theory
could prevent infinite production [27].  In the context of the 2-dim gravity
theory we have studied in this paper there is a natural mechanism which
suppresses the rate of production of remnant states in ordinary processes.
In fact an external observer sitting in the asymptotically flat region of
our 4-dim solutions finds impossible to excite the states of the infinite
throat region because, as we have seen in sect. 3, this excitation are
suppressed by an infinite mass gap.

\bigskip
{\bf Acknowledgment}

{\noindent
We thank the MURST for financial support.}
\bigskip
\centerline {\bf References}
\smallskip
\halign{#\quad&#\hfil\cr
[1]&D. Garfinkle, G.T. Horowitz and A. Strominger, \PRD {\bf D43},
 3140 (1991);\cr
[2]&G.W. Gibbons and K. Maeda, \NPB {\bf B298}, 741 (1988);\cr
 &A. Shapere, S. Trivedi and F. Wilczek, \MPLA {\bf A6}, 2677 (1991);\cr
 &A. Sen, \PRL {\bf 69}, 1006 (1991);\cr
[3]&M. Cadoni and S. Mignemi, in press on \PRD {\bf D48};\cr
[4]&C.F.E. Holzhey and F. Wilczek, \NPB {\bf B380}, 447 (1992);\cr
[5]&J. Preskill, P. Schwarz, A. Shapere, S. Trivedi and F. Wilczek, \MPLA\cr
   &{\bf A6}, 2353 (1991)\cr
[6]&S. Mignemi and N.R. Stewart, \PRD {\bf D47}, 5259 (1993);\cr
[7]&J. Preskill, Preprint CALT-68-1819;\cr
 &S.B. Giddings, Preprint UCSB TH-92-36;\cr
[8]&S.B. Giddings and A. Strominger, \PRD {\bf D46}, 627 (1992);\cr
[9]&C.G. Callan, S.B. Giddings, J.A. Harvey and A. Strominger, \PRD {\bf D45},
\cr &1005 (1992);\cr
[10]&T. Banks, A. Dabholkar,  M. R. Douglas and M. O'Loughlin, \PRD {\bf D45},
\cr &3607 (1992);\cr
 &L. Susskind and L. Thorlacius, \NPB {\bf B382}, 123 (1992);\cr
 &S.P. de Alwis, \PRD {\bf D46}, 5429 (1992);\cr
[11]&S. Trivedi, \PRD {\bf D47}, 4233 (1993);\cr
 &P.-J. Yi, \PRD {\bf D48}, 2777 (1993);\cr
[12]&M. Cveti\v c, A.A. Tseytlin, Preprint CERN TH-6911/93;\cr
[13]&V. Kaplunowsky, \NPB {\bf B307}, 145 (1988);\cr
 &J. Dixon, V. Kaplunowsky, J. Louis, \NPB {\bf B355}, 649 (1991);\cr
[14]&C.G. Callan, J.A. Harvey, A. Strominger, \NPB {\bf B359}, 611
(1991);\cr
[15]& M. Cadoni and S. Mignemi, in preparation;\cr
[16]&R.B. Mann, A. Shiekh and I. Tarasov, \NPB {\bf B341}, 134 (1990);\cr
 &R.B. Mann, Gen. Rel. Grav. {\bf 24}, 433 (1992);\cr
 &Y. Frolov, \PRD {\bf D46}, 5383 (1992);\cr
[17]&G. Mandal, A.M. Sengupta and S.R. Wadia, \MPLA  {\bf A6}, 1685
(1991);\cr
&E. Witten, \PRD {\bf D44}, 314 (1991);\cr
[18]& C. Teitelboim, in {\sl Quantum Theory of gravity }, S.M. Christensen,
ed. (Adam Hilger,\cr & Bristol, 1984); R. Jackiw, {\sl ibidem};\cr
&C. Teitelboim, Phys. Lett. {\bf 126B}, 41 (1983);\cr
&R. Jackiw, \NPB {\bf B252}, 343 (1985);\cr
[19]& S.W. Hawking and G.F.R. Ellis, {\sl The large scale structure
 of spacetime} (Cambridge \cr&Un. Press, 1973);\cr
[20]&J.P.L. Lemos and P.M. S\'a, Preprint DF/IST-8.93;\cr
[21]&G. Dzhordzhadze, A. Pogrelkov and M. Polianov, Sov. Phis. Dokl. {\bf 23},
 828 (1978);\cr
 &Theor. Math. Phys. {\bf 40}, 706 (1979);\cr
 &J. Jackiw, Lectures given at the {\sl International Colloquium on Group
Theoretical}\cr&{\sl Methods in Physics}, Salamanca, June 1992;\cr
[22]&S.J. Avis, C.J. Isham and D. Storey, \PRD {\bf D18}, 3565 (1978);\cr
[23]&S.M. Christensen and  S.A. Fulling, \PRD {\bf D15}, 2088 (1977);\cr
[24]&S. Deser, M.J. Duff and C.J. Isham, \NPB {\bf B11}, 45 (1976);\cr
&P.C.W. Davies, S.A. Fulling and W.G. Unruh, \PRD {\bf D13}, 2720 (1976);\cr
[25]& N.D. Birrell and P.C.W. Davies, {\sl Quantum fields in curved space}
(Cambridge Un.\cr &Press, 1982);\cr
[26]&S.W. Hawking, \PRL {\bf 69}, 406 (1992);\cr
[27]&T. Banks, M. O'Loughlin and A. Strominger, \PRD {\bf D47}, 4476.\cr}

\end